\def\aj{AJ}                   
\def\araa{ARA\&A}             
\def\apj{ApJ}                 
\def\apjl{ApJ}                
\def\ao{Appl.~Opt.}           
\def\aap{A\&A}                
\def\mnras{MNRAS}             
\def\pasp{PASP}               
\def\nat{Nature}              

          [2001/04/25 1.1 (PWD)]
\documentclass[a4paper,twocolumn]{esapub} 
\usepackage{natbib,psfig}

\title{Astrophysical imaging with the Darwin IR interferometer}
\author[1]{H. J. A. R\"ottgering }
\affil[1]{Sterrewacht Leiden, P.O. Box 9513, 2300 RA Leiden, The
Netherlands}
\author[1,2]{L. d'Arcio}
\affil[2]{Space Research Organisation Netherlands, The
Netherlands}
\author[1,3,4]{C. Eiroa}
\affil[3]{Universidad Aut\'onoma de Madrid, Spain} \affil[4]{ESTEC
Noordwijk, The Netherlands}
\author[1]{I. Labb\'e}
\author[5]{G. Rudnick}
\affil[5]{MPA, Garching, Germany}

\begin{document}


\maketitle

\begin{abstract}
The proposed infrared space interferometry mission Darwin has two
main aims: (i) to detect and characterize exo-planets similar to
the Earth, and (ii) to carry out astrophysical imaging in the wavelength 
range 6 - 20 micron at a sensitivity  similar to JWST, but at an angular 
resolution  up to 100 times higher.  In this contribution we
will first briefly discuss the imaging performance of the Darwin
mission.  We will then discuss how Darwin will contribute in a
very significant way to our understanding of the formation and
evolution of planets, stars, galaxies, and super-massive
black-holes located at the centers of galaxies.
\end{abstract}

\section{Introduction}

The main focus of this conference is to discuss the wonderful
recent results on planets orbiting other stars and the prospects
of the Darwin and TPF missions for detecting and characterizing
planets with properties similar to our beloved Earth. For a long
list of reasons, detecting and studying exo-earth's is technically
very challenging. Let us  consider an exo-earth in the habitable zone
of a G-type star at a distance of 10 pc. It will be very dim: in
the IR, it will have a flux density of 0.34 $\mu$Jy (Beichman et al.
1999), \nocite{bei99}  and in the optical magnitude of
$V=30$.  Furthermore, it will be located at a close angular
separation of 100 mas from the parent star.  To be able to detect
and study such earth-like planets very sensitive observations at
high angular resolution are clearly needed. An instrument that is
capable of delivering such exquisite measurements will be able to
also do very important observations for general astrophysics. 
In this contribution 
we will concentrate on the promise of Darwin for
such studies. In the wavelength region around 10 micron, Darwin
will have a similar sensitivity to 
JWST, but with an angular
resolution a factor of $10-100$ larger.

The Darwin mission was an integral part of ESA's Horizon 2000+
programme and was designed from the beginning with these two goals
in mind. The proposal by L\'eger et al. (1996) \nocite{leg96} to
build Darwin originated from a  time before the first exo-planet
was discovered. With more than 100 planets now discovered, the
landscape of this field has changed dramatically, making it even
more clear that a machine that can study exo-earth's is 
needed. Also for ``normal'' astrophysics the scene has changed
significantly, mainly due to the advent of 10-m class optical/near
IR telescopes and X-ray and infrared satellites. 

There are a number of very good reasons to have a Darwin mission
with astrophysical observations as one of its key drivers. First,
it is a logical next step after a long series of IR missions
(IRAS, ISO, SIRTF and JWST). The angular resolution that will be
needed as a follow-up of sources studied by SIRTF and JWST can not
naturally be obained with a single dish telescope, but will
require interferometry.  Second, an important argument is that of
``cost effectiveness'': the Darwin mission will be very expensive,
and if it can do ``two missions at (more or less) the price
of one'', this will be very valuable. Third, there is the
argument of redundancy. The whole mission will be very
challenging, and having more than one goal will help reducing the
scientific and technical risk of the entire mission.

In this presentation, we will first discuss the performance of the
Darwin imaging mission. Subsequently we will briefly describe the
research into three areas of astrophysics (star formation,
galaxy formation and evolution, and active galactic nuclei) that
Darwin will contribute to in a major way.

\section{Imaging considerations}

\subsection{Sensitivity}

The proposed configuration for Darwin consists of 6 telescopes
each with a diameter of 1.5 m and a central beam-combiner. The
relevant systems at the telescopes and beam-combiner will be
passively cooled to 40 K, so that the sensitivity will be limited
by shot-noise from emission by the zodiacal background
(Fridlund et al. 2001, Nakajima and Matshura 2001).
\nocite{fri00,nak01} This system should be a able to detect a
point source of 2.5 $\mu$Jy with a signal to noise of 5 in one hour
of integration time. For sources that are more complex, assessing
the integration time depends on the details of the morphologies. A
simple rule of thumb that is used in the radio astronomical
community, is that an object can be imaged in a certain
integration time provided that the total signal within the primary
field of view is larger by about a factor $50 - 100$ than the
expected off-axis rms noise in the final map. For Darwin the
point-spread function (PSF) for a single telescope is 1.4 arc-sec
at 10 micron. This is the maximum field of view (FOV) if the beam
combination is done in the pupil-plane. The image requirements for
Darwin therefore are that a source can be ``mapped'' with an
integration of one hour, provided that the total flux within $\sim
1$ arc-sec is more than $25 - 50 \mu$Jy.

\subsection{Resolution and filling of the UV plane}

The maximum baselines that are foreseen are about 500 meter, which
translates into a maximum resolution of 4 mas at 10 micron. This
is a factor of more than 100 higher than the nominal resolution at
10 micron of 350 mas of the 6.4 m mirror of JWST. Let us consider
an experiment with baselines of up to 150 meters. With the
resulting resolution and  the FOV as determined by the PSF's of the
individual telescopes, a final map will contain up to $100 \times 100$
individual pixels. This number of pixels is directly set by the
ratio of length of the maximum baseline to  the size of the FOV.
For mapping very complex objects it is essential that the number of
visibilities that are measured 
in the UV plane is significantly
greater than the number of parameters that is needed to describe
the image. Furthermore, these visibility measurements should be
well distributed over the UV plane. This means that in the case
that our complex object occupies a significant fraction of the
100$^2$ pixels, also on  the order of 100$^2$ visibility measurements 
will be needed.

\begin{figure}[tb]
\centerline{
\psfig{figure=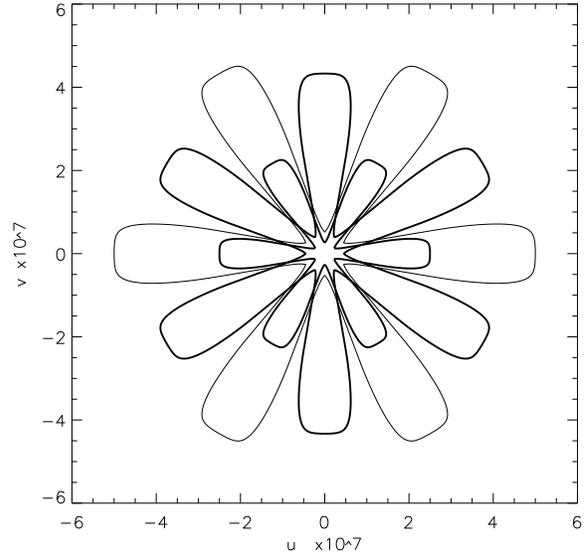,width=8cm}}
\caption{\label{uv} The coverage of the UV plane for the 
single expansion of the configuration to maximum 
baselines of 500 meter coupled with a 60 degree rotation.
From d'Arcio et al. 2003.}
\end{figure}

Per Darwin configuration, a maximum of 15 independent visibility
points can be obtained. This immediately indicates that a 100$^2$
map would need visibility measurements from 
666 different configurations. This is not such a
severe requirement if we take a closer look at the dynamics of the
configuration. For a more detailed discussion, see d'Arcio et al.
2003. \nocite{dar03} A basic approach to filling up the UV plane
might be a single expansion coupled with a 60 degree rotation
For the resulting UV-plane coverage, see Fig. \ref{uv}. 
With the planned  thrusters delivering 1 mN, the fastest
reconfiguration cycle takes about 16 hours. The average 
baseline 
rate will be about 1.5 cm s$^{-1}$.  The maximum integration time
for a single UV point is then set by the requirement that each UV
point should not be significantly blurred by the movement of the
array. For a telescope of 1.5 meter a movement of 15 cm in the UV
plane is certainly tolerable, indicating maximum individual
integration times of about 10 seconds. This would yield 15 * 16
hours
* 3600 sec / 10 sec = 86400 different UV samples, far  more than
our simple required minimum of 666 samples needed for complex
maps.

Interestingly, this leads to the conclusion that the free flying
telescopes allow for much ``easier'' filling of the UV plane than
is possible with fixed ground based telescopes. Furthermore, the
number of UV points needed is modest compared to conventional
radio synthesis telescopes.  For example, for the VLA the ratio of
maximum baseline size over the size of the individual apertures is
27 km/25 m = 1000, a factor of 10 more than Darwin. For very large
baseline interferometry with radio telescopes scattered all over
the globe, the ratio is many orders of magnitude worse.

\subsection{Co-phasing}

While the telescopes are moving it is essential that the system is
able to integrate to obtain the desired visibility measurements.
Note that since the aim is to perform imaging, both the amplitude
and the phase of the visibilities need to be measured.  To be able
to do this, the array needs to remain co-phased while moving. In
general, the science targets are too faint to give enough signal
for co-phasing. Instead the light from an off-axis bright reference
star is used. A system based on such a principle is currently
being implemented at the VLTI under the name PRIMA, which stands
for Phase-Referenced Imaging and Microarcsecond Astrometry.  For a
detailed description of this instrument we refer to the
contribution of Quirrenbach to these proceedings.

d'Arcio et al. 2003 \nocite{dar03} discussed the various options for
implementing off-axis referencing.  An important design choice is
related to how to multiplex the science and reference beam.  Sending
those two beams along two different path is complicated: it requires
relatively large optics and the calibration of the differential path
lengths is arduous. Sending the two beams along a common path is
therefore clearly preferred. Multiplexing the beams can in principle
be done using polarization or splitting up into time
sequences. However, multiplexing in wavelengths seems simplest and has
therefore been selected by the Alcatel study (Fridlund et
al. 2000). Within such a scheme the $K$-band is used for
fringe-tracking. At the expected limiting magnitude for fringe
tracking of m$_k$=17, almost 100 \% of the sky should be observable,
provided that the angular distance of the science target and the
reference star is allowed to be maximally 1.5 arcminute. An issue is
cross talk between the science and reference channels, especially at
the shorter science-wavelengths. d'Arcio et al. (2003) showed that
levels of a few per cent cross-talk can be tolerable and this is
probably achievable with a proper design.

The implementation of such a scheme studied by Alcatel involved 
modifying the nulling beam combiner. This however, strongly reduces
the sensitivity, potentially by a factor of about $10 - 20$
(d'Arcio et al. 2003). For imaging it would therefore clearly
be advantageous to have a separate beamcombiner.

\begin{figure*}[th]
\centerline{
\psfig{figure=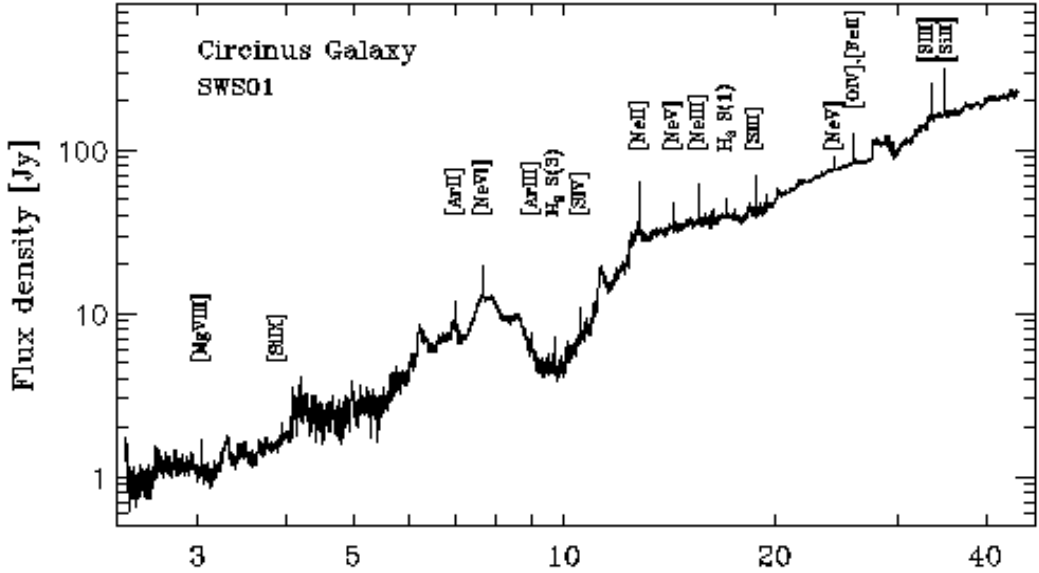,width=14cm,clip=} }
\caption{\label{circ} 
The infrared spectrum of the 
famous starburst galaxy Circinus as observed by ISO (from Moorwood 
et al. 1996) 
} 
\end{figure*}

\subsection{Extending the field of view}

In the case of pupil-plane combination, the field of view of
Darwin will be small, of the order of 1 arc-sec. Mapping a larger
field of view can always be done through a mosaic of different
pointings. This is of course very expensive in time.  A 
natural way of enlarging the field of view is to combine the beams
in the image plane. This is somewhat complex since the pupils need
to be remapped at the beam-combiner with the correct
demagnification and relative distances. Furthermore, the required 
optics are large in size.  An interesting idea, which potentially
simplifies the design is to have zoom optics installed at the
telescopes, so that at the level of the beam-combiner, the pupils
do not need to be reconfigured.  In the implementation of such a
design that d'Arcio and le Poole discuss in these proceedings, a
4k*4k detector would yield a variable field of view, ranging from
25 arcsec for baselines of 50 m down to 2.5 arcsec for baselines of
500 m.

\section{Science with Darwin}

\subsection{Diagnostics}

The wavelength region between 5 and 20 micron is very rich in
diagnostics for a number of species, including molecules, ions, dust
and late type stars (e.g. van Dishoeck 2000). \nocite{dis00} As a nice
illustration of this richness, we show in Fig. \ref{circ} the infrared
spectrum of the famous starburst galaxy Circinus as observed by ISO
from Moorwood et al. (1996).

For molecules, rotational and vibrational lines can be observed
and for ions, forbidden fine-structure lines. Combined, these can
be used to map out the temperature, density, and metallicity,
kinematics of gas at intermediate to cold temperatures ($10 -
10,000 $ K) and moderate densities $10 - 10^6$ cm$^{-3}$. 
In this wavelength regime 
the continuum is often due to emission from dust and
Polycyclic Aromatic Hydrocarbon (PAH) molecules. Studying
the precise form of the continuum reveals information about the
composition and temperature structure of the various dust and PAH
components.  The spectral energy distribution of a normal nearby
galaxy peaks at a rest-frame wavelength of a few microns. For
galaxies at high redshifts, this peak shifts to observed
wavelengths of $5-10$ microns. High resolution sensitive
measurements at $5- 10$ microns are therefore very important for
understanding the formation and evolution of high redshift
galaxies.

\section{Star formation}

Based on a wealth of observations a sketch of a scenario showing how (isolated
low-mass) stars form seems to be in place (Shu et
al. 1987). \nocite{shu87} Also the suggested classification scheme for 
the various evolutionary stages seems commonly accepted.  The
formation of a star commences with the growth of condensations in
molecular clouds. When the density of a condensation reaches a
critical value, a collapse sets in, on a time scale likely to be
governed by the local sound speed.  Subsequently, a protostar
surrounded by an accretion disk forms, both deeply embedded within an
envelope of infalling dust and gas.  A striking phenomenon
during this phase is the occurrence of bipolar outflows.  Gradually,
the inflowing matter will fall more and more onto the disk rather than
the star. Within the final stage, the disk
might be fully dispersed, for example 
by an energetic outflow. Alternatively, the
disk material might (partially) coagulate and form one or more
planets.

Within the vast range of physical conditions that occur during
star formation, the high resolution imaging provided by Darwin
will map the density, temperature, metallicity and dynamics of
young stellar objects. This is particular relevant for studies of
the core inflow, the accretion disks, the location of the jet
outflow and planet formation. For example, it is not well
understood how the jet outflow originates from the accretion disk.
How does this start? What keeps the jet stable? Mapping out
the structure of the disc will be very important for
constraining models of planet formation.

For a first impression of how many sources are available
for detailed study 
with Darwin, we have made a compilation of various ISOCAM
surveys at 6.7 and 14.3 micron down to a completeness limit of $\sim
10-15$ mJy of three starforming regions (Persi et al. 2000, Bontemps
et al. 2001, Kaas, priv. comm.). \nocite{per00}
In Table 1, we give for each of these regions
their names, the estimated age of the systems, their distance and the
total  area surveyed.  Furthermore, the total number of sources
detected both at 6.6 and 14.3 micron is given and, if possible, the
number of sources in each of the star formation classes I, II and
III. This classification is done on the basis of the 6.7 and 14.3
colour indices.  For more details see Bontemps et al. 2001.
\nocite{bon01}

\begin{table}
  \begin{center}
    \caption{Results from a compilation of 6.7 and 14.3 micron
    ISOCAM surveys. 
    }\vspace{1em}
    \renewcommand{\arraystretch}{1.2}
    \begin{tabular}[h]{l|ccc}
      \hline
Region                     & Chameleon & Ophiucus & Serpens \\
\hline
Age (Myr)                  &  5        &  0.2-3   & 0.1 \\
Distance (pc)              & 160       &  160     & 260 \\
Observed Size              & 0.57      &  0.7     & 0.13 \\
\qquad (deg. sq.) \\
N(6.6 + 14.3)              &  103      &  211     & 124 \\
N(Class I)                 &           &  16      & 19 \\
N(Class II)                &  46       & 123      & 44 \\
N(Class III)               &  19       &  77     &  9 \\
      \hline
      \end{tabular}
    \label{tab:table}
  \end{center}
\end{table}

\begin{figure*}[tb]
\centerline{
\psfig{figure=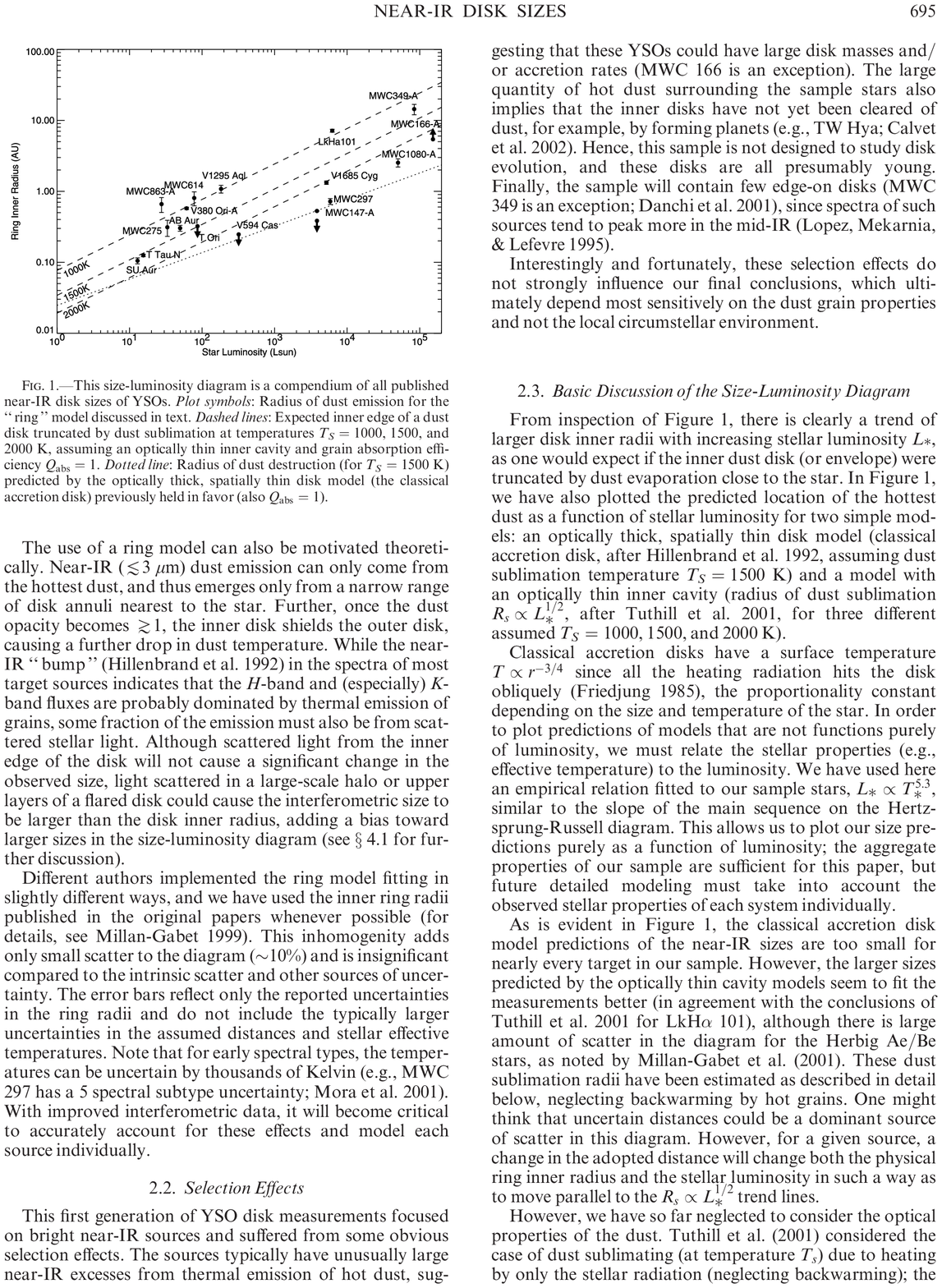,width=12cm,clip=}}
\caption{\label{mg} 
The scale sizes at 1.6 and/or 2.2 $\mu$m as determined from
interferometric measurements of circumstellar disks around young
stellar objects (YSOs), mainly Herbig Ae/Be and T Tauri stars versus
the luminosity of the parent star (from Monnier and Millan-Gabet 2002).} 
\end{figure*}

To investigate whether these sources have the appropriate angular
scale for observations with Darwin, we use as a starting point the
work of Monnier and Millan-Gabet (2002). \nocite{mon02} They have
compiled scale sizes at 1.6 and/or 2.2 $\mu$m for the
circumstellar disks around young stellar objects (YSOs), mainly
Herbig Ae/Be and T Tauri stars. They find that the inner radii of
the disks range from 0.1 AU up to 10 AU and appear to be correlated
with the luminosity of the star (see Fig. \ref{mg}). If we now
apply this rough correlation between luminosity and size to the
Class II sources as presented in Tabel 1., we
find that the majority of these sources have
estimated angular sizes between 10 and 500 mas.  An alternative
way to estimate characteristic angular sizes for such objects is
to use models of proto-planetary disks. A family of such models
have been developed by D'Alessio et al. (2001). \nocite{dal01}
These models solve the radiative transfer through a viscous dusty
disk heated by the central region.  The resulting synthetic SEDs
can then be compared to observed SEDs. Mer\'\i n et al.
(these proceedings)  used these models to predict angular size scales for
proto-planetary disks and find angular size scales in the range of
ten to a few hundred mas, consistent with the observationally based
estimate for the sources presented in Table 1.

From these two estimates it is clear that a proper population
study requires the sensitivity and angular resolution of the Darwin mission.

\section{Active galaxies}

Active galaxies are among the most spectacular objects in the sky.
Their compact nuclei can be so luminous that they can outshine an
entire galaxy.  These objects show many interesting observational
characteristics (e.g. radio lobes, jets, narrow and broad emission
lines, polarized light, X-ray continuum and line emission,
see reviews by Antonucci 1993 and Urry and Padovani 1995).
\nocite{ant93,urr95} The relative importance of these
observational characteristics varies dramatically from object to
object, and a complicated phenomenological scheme has emerged to
classify objects into many different types of AGN, including Sey1,
Sey2, QSOs, QSRs, BLACs, radio galaxies, etc.  A major issue in
this field is how to understand all these different characterics
on the basis of the fundamental physical and geometrical
properties of the inner heart of the AGN.  Orientation is an
important factor in determining the appearance of an individual
object.  Furthermore, the characteristics of the central massive
black hole (including total mass and spin) are likely to be
directly related to the total output of the AGN and the presence
of well defined jets. And finally, over the lifetime of the AGN
activity its appearance can change dramatically (for example radio
sources can grow up to a few Mpc, starting at the pc scale).  This
evolution of AGNs is also evident in the fact that at a redshift
of $z=2$ bright AGNs are 1000 times more frequent than today.
Although it is clear that this indicates a close link between the
formation of galaxies and the AGN phenomenon the reason for the
existence of this AGN epoch is not understood. More direct
evidence for this link between AGNs and galaxy formation is
provided by the relationship between galaxy and massive black hole
mass, increasing the significance of AGNs for  the overall
understanding of galaxy formation.

In the currently popular and attractive ``unified''
model of active galactic nuclei, all AGNs contain the
following components: A central black hole fed by an accretion
disk, which is surrounded by an optically thick obscuring torus and
broad and narrow line regions.  The orientation
with respect to the line of sight determines whether the object is
observed to have  broad emission lines, originating
from within the hot central hole of the torus, or whether the torus
blocks this region from view, leaving only the unobscured narrow line
region visible.

Although this picture is capable of explaining a large number of the
differences between the various classes of AGN (e.g. Antonucci 1993,
Urry \& Padovani 1995), \nocite{ant93,urr95}
it is still a vigorous debate whether other
mechanisms contribute to, or even dominate over, the scenario in which
orientation and a putative torus play such a major role. It has even
been argued that in a subset of AGNs, the main power-source is not the
black-hole but supernova explosions produced within a central
starburst region. It is not clear, however, if all AGNs have starburst
regions, whether all starburst galaxies contain AGNs nor what the
causal relation is between these two phenomena.

An important issues in understanding AGN unification are therefore
whether dust tori exist and whether the physics/geometry of such tori
can be constrained well enough for many of the differences between
AGNs to be understood.

A number of radiative transfer models for obscuring tori models have
been proposed (e.g.  Krolik and Begelman 1988; Pier and Krolik 1992;
Efstathiou and Rowin-Robinson 1995; Granato et al. 1997; Manske et al.
1998; Wolf and Henning 1999).
\nocite{kro88,pie92,efs95,gra97,man98,wol99}   Within these models the
radiative transfer through a torus with a given geometry, optical
depth and dust grain composition is calculated. The model subsequently
gives the emerging IR spectrum and the morphology as a function of
wavelengths.

\begin{figure}[tb]
\centerline{
\psfig{file=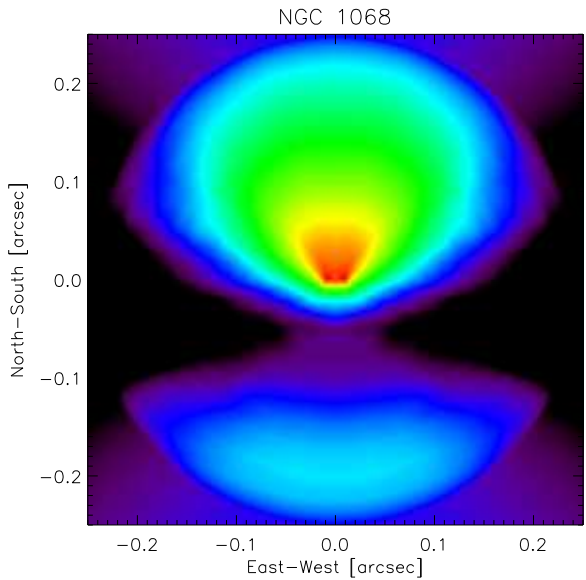,width=8cm} }
\caption{\label{torusmodel} Greyscale representation of the
morphology at 10 micron according to the model for NGC 1068 of
Granato et al. (1997). For this particular model the 
visual extinction is $A_V=$ 150 mag
and the orientations is theta = 60$^\circ$
(Courtesy Bjon Heijligers) } 
\end{figure}

It is very clear that Darwin will be able to map
nearby tori in exquisite detail. Recently the famous Seyfert 2
galaxy NGC 1068 was observed at the VLT interferometer using the
10 micron MIDI instrument (see ESO press release 17/03 at
http://www.eso.org/outreach/press-rel/pr-2003/pr-17-03.html). This
first detection of fringes of an extragalactic object is an
important step for optical interferometry. The plan is that during
2003 of the order of $10 - 20$ visibility points will be measured.
This will give important constraints on the general shape of the
NGC 1068 torus. However, compared to VLTI, Darwin will  be a
dramatic step forward. The number of visibilities that will be
obtained will be 3 orders of magnitude higher, with a fairly uniform
filling of the UV-plane out to 500 meter. This will be very
important to map in detail substructure, warps, and dynamics of
nearby tori.

\begin{figure}[tb]
\centerline{
\psfig{figure=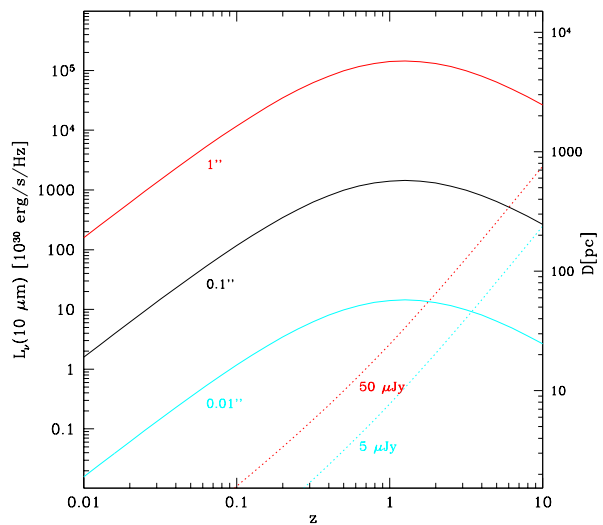,width=8cm} }
\caption{\label{agn} The 10 micron luminosity and physical scale D of
dusty tori as a function of redshift.  The solid lines
are for observed angular scales
of 1, 0.1 and 0.01 arcsec.  The dotted lines are for
an observed  10 micron flux density
of 5 and 50 $\mu$Jy.  The computations have been done for $H_0 = 75$ km
s$^{-1}$ Mpc$^{-1}$ and $\Omega=1$. }
\end{figure}

An interesting question is to what distance tori of AGN can be
mapped. To investigate this, we will use the torus models 
of Granato et al. 1997.  In these models, the inner radii of these
tori, $r_{\rm in}$, are set by the distance from the central source
at which the dust grains sublimate due to the strong nuclear
radiation. This radius is larger for more luminous AGN. For the
models of Granato et al., $r_{\rm in} \sim 0.5 L_{46}^{1/2}$ pc,
where $L_{46}$ is the luminosity of the central optical UV emitter
in units of $10^{46}$ ergs s$^{-1}$.  As a scale size of the torus
D we will use 300 $r_{\rm in}$.  In Figure \ref{agn}, the diameter
D and the 10 micron luminosity (which are directly coupled) are
given as a function of $z$ for angular scales of 1, 0.1 and 0.01
arcsec.  The dotted lines correspond to the 10 micron luminosities
as a function of $z$ for 10 micron flux densities of 5 and 50
$\mu$Jy.

A relatively weak AGN such as NGC 1068 has a 10 micron luminosity of
the order of $1.7 \times 10^{31}$ erg s$^{-1}$ Hz$^{-1}$ and its
modelled torus size is 60 pc. Up to redshifts of $z=1-2$ such weak AGN
are brighter than the nominal sensitivity for a one hour imaging
observation (see Section 2.) of $25-50 \mu$Jy. Also the
nominal resolution at 10 micron of 20 mas is very adequate for imaging
the tori at these redshifts.  Brighter AGN can basically be mapped up
to redshift of $z=10-20$ (if they exist).

This shows that Darwin can not only study the physics of dusty tori in
our local universe, but also at large redshifts.  This will give the
unique opportunity to investigate how the properties of tori change
with redshift and when and how these tori and their associated massive
black-holes are built up at an epoch when galaxies are forming.

\section{Galaxy formation and Evolution }

With the advent of 10-m telescopes, it is now possible to define
and study large samples of very distant galaxies. This work has
been pioneered by Steidel and coworkers. Using the ``Lyman break
technique'', they have defined a sample containing of order a 1000
galaxies between $2.5 < z < 5$ (e.g. Steidel et al. 1999).
\nocite{ste99a} Other techniques to obtain samples of very distant
galaxies are to carry out deep imaging using  narrow band filters
to find Ly$\alpha$ and H$\alpha$ emitting galaxies (e.g. Venemans
et al. 2002; \nocite{ven02} Kurk 2003) or preselect on very red
$J-K$ colours (Franx et al. 2003).  \nocite{kur03,fra03} Also
direct spectroscopic follow-up of SCUBA galaxies (e.g. Chapman et
al. 2003), \nocite{cha03} radio galaxies (e.g. de Breuck et al.
2001) \nocite{bre00a} and X-ray emitters (e.g. Rosati et al. 2002)
\nocite{ros02b} can yield significant samples of $z>2$ objects.

This amazing observational progress of the last 10 years has been
accompanied by a large body of work to build reliable and robust
models of galaxy formation.  The goal is to model the evolution of
galaxies with, as the main input, the physical conditions as they
existed in the very early universe. For a thorough account we
refer to a few of the excellent articles and reviews that have
been written (see for example: Rees and Ostriker 1997; White and
Rees 1978; White and Frenk 1991; Cole 1991; Lacey and Silk 1991;
Kauffmann et al. 1993; Lacey et al. 1993; Cole et al. 1994;
Kauffman et al. 1994; Somerville et al. 2001; Moustakas and
Sommerville 2002; Bell et al. 2003)

\nocite{ree77,whi78, whi91, col91, lac91, kau93, lac93c, col94,
kau94, som01, bel03, mou02}

\begin{figure*}[t]
\centerline{ \psfig{figure=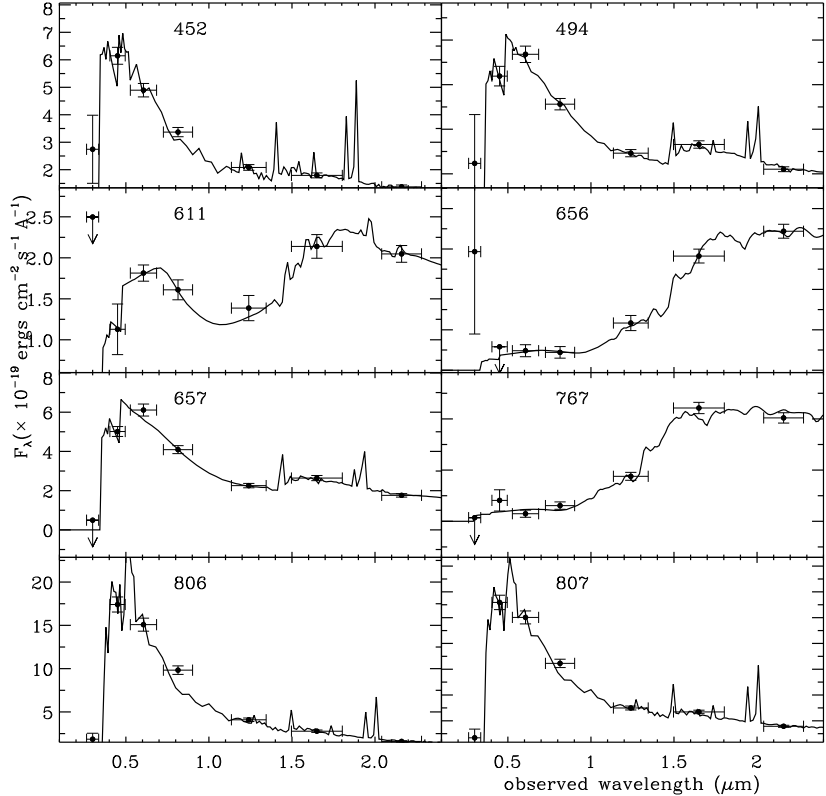,width=12cm} }
\caption{\label{fits} Sample of template fits to photometric data
for 8 objects in the HDF-S. The photometric data are from the HDF-S 
observations and from the FIRES project. The derived template
fitting yielded photometric redshifts  between $2.3 < z< 3.6$. For
further information see Labb\'e et al. (2003).  }
\end{figure*}

Virtually all of the current models assume that the dynamics of
the large scale mass distribution in the early universe is driven
by the gravity exerted by some form of dark matter.  At early
times the density fluctuations within this medium are  modelled by
a described distribution whose functional form depends on the
physics in the early universe, the nature of the dark matter and
a suitable choice of cosmological parameters.  The evolution of
the dark matter distribution can be studied analytically or with
the help of N-body simulations.  A second step is to include the
baryonic gas and to follow its hydrodynamic evolution. Gas
dynamics, shocks and radiative heating and cooling all need to be
part of the simulation to obtain a realistic multi-phase medium.
The outcome of this kind of simulation is that a significant
fraction of the gas cools and settles at the centers of dark
matter halos.  It is from that gas that the stars that will make
up future galaxies will form.  Often it is then simply assumed
that the rate at which the gas at the center of these halos forms
stars is proportional to the total amount of gas present and
inversely proportional to the dynamical timescale within the dark
matter halo. The rate with which these proto-galaxies forms stars
is limited due to both supernovae and stellar winds which blow gas
out of the centers of these halos.  The merging of galaxies
greatly enhances the combined star formation rate of both
galaxies, possibly to a rate at which a very large fraction of the
gas is transformed into stars within a few dynamical timescales.
Finally, the combination of the inferred star formation rate, an
assumed initial mass function and the spectral evolution of
individual stars will then give the evolution of the integrated
spectra of an individual galaxy.  With this kind of modelling
gross properties of the general galaxy population can be
calculated.  These properties include the luminosity function of
galaxies, the redshift distribution, the relative numbers of
ellipticals and spirals, faint galaxy counts, the history of star
formation. Also more detailed properties like the size
distribution, and the Tully-Fisher relation can be predicted with
a fair accuracy.

What role can Darwin in this area?  Maybe it is important
to realise that even 10 years from now the issue of how galaxies form
will still be one of the most important topics within the field of
astronomy. Or to quote Rees (1998), \nocite{ree98}
who states that in 10 years from now, ``we will probably
still be unable to compute crucial things like the star formation
efficiency, feedback from supernovae. etc -- processes that current
models for galactic evolution are forced to parametrise in a rather ad
hoc way''.

An essential constraint on these galaxy formation models will be 
the spatial structure of very distant galaxies at $1-2$ micron
restframe, the location of the peak of the spectral energy
distribution of nearby galaxies. For $z\sim 5$ galaxies, this region
is redshifted into the spectral window within which Darwin will be
observing.

A crucial question is  whether there are sufficient number of distant
galaxies on the sky with appropriate angular sizes for Darwin to observe. 
To investigate this we will use results from the Faint InfraRed
Extragalactic Survey (FIRES, Franx et al. 2000),  which
is a very deep infrared survey centered on the Hubble Deep Field South
using the ISAAC instrument mounted on the VLT (Moorwood 1997).
\nocite{fra00a,moo97b} With integration times of more than 33 hours for
each of the infrared bands $J$, $H$ and $K$, limiting AB magnitudes of
26.0, 24.9, and  24.5 respectivily are reached (Labb\'e et al. 2003).
\nocite{lab03}  A major advantage of observing this field is that
multicolour HST photometry is available.

Using the multi color data, we estimate the distance of the objects
using the photometric redshift technique.  This technique uses a fit to
the observed SED with a linear combination of galaxy templates. The
procedure is described in detail in Rudnick et al. 2001. and 2003.
\nocite{rud01,rud03}  
An example of fits that are obtained is given in Fig. \ref{fits}. The
galaxies shown are a mixture of red J-K galaxies and Lyman Break
galaxies all at a photometric redshift of $z\sim3$. For further
details, we refer to Labb\'e et al. 2003.

\begin{figure}[tb]
\centerline{ \psfig{figure=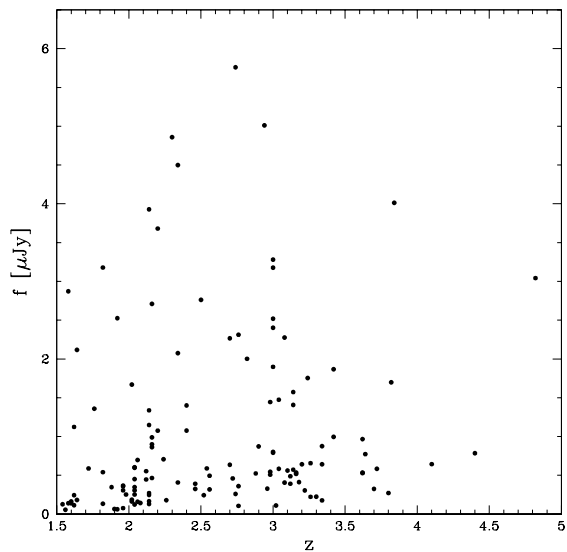,width=8cm} }
\caption{\label{flux} 10 micron flux densities estimated from the
template fits for a subsample of the FIRES survey with $K_{AB} > 25$
and $1.5 < z_{\rm phot} < 5$.}
\end{figure}

For these high redshifts, the reddest observed colour, $K$-band, samples only the
rest-frame $V$-band. To sample the SED in the restframe $K$-band
observations at the Darwin wavelength range are essential. This is
important so that a complete census of the stellar population in
these distant galaxies  can be obtained, not biased due to dust
obscuration or short bursts of massive star formation. To estimate
the observed 10 micron flux densities, we used the fits to the HST
and FIRES photometry. In Fig. \ref{flux} we show the results from a sample
with $K_{AB} > 25$ and $1.5 < z_{phot} < 5$. The $K$-band selection was
chosen such  the the photometric redshifts are fairly reliable (see
Labb\'e et al. 2003 for details). The lower limit to 
the photometric redshift was
chosen so that, at an observed wavelength of 10 micron, the SED is
still dominated by stellar light and not by a hot dust component.
From Fig \ref{flux}, we conclude that, for the brighter objects, it is
possible to obtain good images with signal to noise ratios of 50
within integration times of 25 - 50 hours (see Section 2.).

\begin{figure}[tb]
\centerline{ \psfig{figure=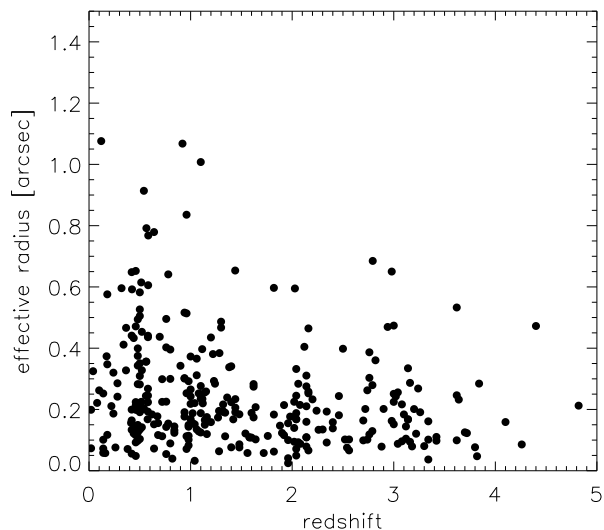,width=9cm} }
\caption{\label{size} The effective radii as measured 
in the $K$-band as a function of redshift for 
a subsample of the FIRES survey with $K_{AB} > 25$} 
\end{figure}

The second important issue is related to the angular size scales of these
FIRES objects. In Fig. \ref{size}, we plot the effective radius of the
objects in FIRES sample versus redshift. The overall trend is that
the angular sizes tend to decrease with redshift. Careful
modelling of the various selection effects involved shows that the
physical sizes of luminous galaxies ($L_V>2 \times 10^{10}L_{\rm
sun}$) at $2<z<3$ are 3 times smaller than that of  equally luminous
galaxies today (Trujillo et al. 2003). \nocite{tru03} The size
distribution shows that twice the median effective
radius is similar to the resolution (FWHM) of the JWST or in other
words, JWST will hardly resolve these distant galaxies. It is
clear that for a good study of the morphology of especially the
older stellar population, Darwin will be very important.

In R\"ottgering et al. 2000 \nocite{rot00a} we have used the models
of Devriendt and Guiderdoni (2000)  
to discuss Darwin's ability to detect and study
distant galaxies.  
\nocite{dev00} It was  appropriate to use these
models since they include the extinction and emission of dust and as
such the models correctly fit the ISO 15 $\mu$m source counts.
Although a detailed comparison between these models and the FIRES data
just presented is in progress, the predicted redshift and angular size
distributions also lead to the conclusion that the flux densities and
angular sizes are well in the range of  Darwin's capabilities.

\section{Conclusion}

The scientific potential of the Darwin mission for general astrophysical
imaging is tremendous.  With the present performance of $\mu$Jy
sensitivity and tens of mas angular resolution, it will allow for
detailed studies addressing a number of crucial questions related to the
formation and evolution of planets, stars, AGN and galaxies.

 \section*{Acknowledgments}

I would like to acknowledge the pleasant discussions and important
input from a number of colleagues including Bruno Guiderdoni, Ewine
van Dishoeck, Malcolm Fridlund, Emanuele Daddi, GianLuigi Granato,
Marijn Franx, Bjorn Heijligers, Klaus Meisenheimer, Rudolf Le Poole
and Jan Willem den Herder.


\begin{thebibliography}{47}
\expandafter\ifx\csname natexlab\endcsname\relax\def\natexlab#1{#1}\fi
\expandafter\ifx\csname url\endcsname\relax
  \def\url#1{{\tt #1}}\fi
\expandafter\ifx\csname urlprefix\endcsname\relax\def\urlprefix{URL }\fi

\bibitem[{{Antonucci}(1993)}]{ant93}
{Antonucci} R., 1993, \araa, 31, 473

\bibitem[{Beichman et~al.(1999)Beichman, Woolf, \& Lindensmith}]{bei99}
Beichman C.A., Woolf N.J., Lindensmith C.A. (eds.), 1999, Terrestrial Planet
  Finder, Origin of Stars, Planets and Life, Jet Propulsion Laboratory

\bibitem[{{Bell} et~al.(2003){Bell}, {Baugh}, {Cole}, {Frenk}, \&
  {Lacey}}]{bel03}
{Bell} E.F., {Baugh} C.M., {Cole} S., {Frenk} C.S., {Lacey} C.G., Mar. 2003,
  ArXiv Astrophysics e-prints

\bibitem[{{Bontemps} et~al.(2001){Bontemps}, {Andr{\' e}}, {Kaas}
  et~al.}]{bon01}
{Bontemps} S., {Andr{\' e}} P., {Kaas} A.A., et~al., Jun. 2001, \aap, 372, 173

\bibitem[{{Chapman} et~al.(2003){Chapman}, {Blain}, {Ivison}, \&
  {Smail}}]{cha03}
{Chapman} S.C., {Blain} A.W., {Ivison} R.J., {Smail} I.R., Apr. 2003, \nat,
  422, 695

\bibitem[{{Cole}(1991)}]{col91}
{Cole} S., Jan. 1991, \apj, 367, 45

\bibitem[{{Cole} et~al.(1994){Cole}, {Aragon-Salamanca}, {Frenk}, {Navarro}, \&
  {Zepf}}]{col94}
{Cole} S., {Aragon-Salamanca} A., {Frenk} C.S., {Navarro} J.F., {Zepf} S.E.,
  Dec. 1994, \mnras, 271, 781

\bibitem[{{D'Alessio} et~al.(2001){D'Alessio}, {Calvet}, \& {Hartmann}}]{dal01}
{D'Alessio} P., {Calvet} N., {Hartmann} L., May 2001, \apj, 553, 321

\bibitem[{{D'Arcio} et~al.(2003){D'Arcio}, {den Herder}, {Le Poole}, \& {R{\"
  o}ttgering}}]{dar03}
{D'Arcio} L., {den Herder} J., {Le Poole} R.S., {R{\" o}ttgering} H.J.A., Feb.
  2003, In: Interferometry in Space. Edited by Shao, Michael. Proceedings of
  the SPIE, Volume 4852, pp. 184-195 (2003)., 184--195

\bibitem[{{De Breuck} et~al.(2000){De Breuck}, {R{\"o}ttgering}, {Miley}, {van
  Breugel}, \& {Best}}]{bre00a}
{De Breuck} C., {R{\"o}ttgering} H., {Miley} G., {van Breugel} W., {Best} P.,
  Oct. 2000, \aap, 362, 519

\bibitem[{{Devriendt} \& {Guiderdoni}(2000)}]{dev00}
{Devriendt} J.E.G., {Guiderdoni} B., Nov. 2000, \aap, 363, 851

\bibitem[{{Efstathiou} \& {Rowan-Robinson}(1995)}]{efs95}
{Efstathiou} A., {Rowan-Robinson} M., Apr. 1995, \mnras, 273, 649

\bibitem[{{Franx} et~al.(2000){Franx}, {Moorwood}, {Rix} et~al.}]{fra00a}
{Franx} M., {Moorwood} A., {Rix} H., et~al., 2000, The Messenger, 99, 20

\bibitem[{{Franx} et~al.(2003){Franx}, {Labb{\' e}}, {Rudnick} et~al.}]{fra03}
{Franx} M., {Labb{\' e}} I., {Rudnick} G., et~al., Apr. 2003, \apjl, 587, L79

\bibitem[{{Fridlund et al.}(2000)}]{fri00}
{Fridlund M., et al.} M., 2000, Darwin, the Infrared Space Interferometer,
  Concepts and Feasibility Study Report, ESA report ESA-SCI, 2000, 12

\bibitem[{{Granato} et~al.(1997){Granato}, {Danese}, \& {Franceschini}}]{gra97}
{Granato} G., {Danese} L., {Franceschini} A., Sep. 1997, \apj, 486, 147

\bibitem[{{Kauffmann} et~al.(1993){Kauffmann}, {White}, \&
  {Guiderdoni}}]{kau93}
{Kauffmann} G., {White} S.D.M., {Guiderdoni} B., Sep. 1993, \mnras, 264, 201

\bibitem[{{Kauffmann} et~al.(1994){Kauffmann}, {Guiderdoni}, \&
  {White}}]{kau94}
{Kauffmann} G., {Guiderdoni} B., {White} S.D.M., Apr. 1994, \mnras, 267, 981

\bibitem[{{Krolik} \& {Begelman}(1988)}]{kro88}
{Krolik} J.H., {Begelman} M.C., Jun. 1988, \apj, 329, 702

\bibitem[{{Kurk}(2003)}]{kur03}
{Kurk} J., 2003, The Cluster Environment and Gaseous Halos of Distant Radio
  Galaxies, Ph.D. thesis, , Univ.\ Leiden, (2003)

\bibitem[{{Labb{\' e}} et~al.(2003){Labb{\' e}}, {Franx}, {Rudnick}
  et~al.}]{lab03}
{Labb{\' e}} I., {Franx} M., {Rudnick} G., et~al., Mar. 2003, \aj, 125, 1107

\bibitem[{{Lacey} \& {Silk}(1991)}]{lac91}
{Lacey} C., {Silk} J., Nov. 1991, \apj, 381, 14

\bibitem[{{Lacey} et~al.(1993){Lacey}, {Guiderdoni}, {Rocca-Volmerange}, \&
  {Silk}}]{lac93c}
{Lacey} C., {Guiderdoni} B., {Rocca-Volmerange} B., {Silk} J., Jan. 1993, \apj,
  402, 15

\bibitem[{{Leger} et~al.(1996){Leger}, {Mariotti}, {Mennesson} et~al.}]{leg96}
{Leger} A., {Mariotti} J.M., {Mennesson} B., et~al., Oct. 1996, Icarus, 123,
  249

\bibitem[{{Manske} et~al.(1998){Manske}, {Henning}, \& {Men'shchikov}}]{man98}
{Manske} V., {Henning} T., {Men'shchikov} A.B., Mar. 1998, \aap, 331, 52

\bibitem[{{Monnier} \& {Millan-Gabet}(2002)}]{mon02}
{Monnier} J.D., {Millan-Gabet} R., Nov. 2002, \apj, 579, 694

\bibitem[{{Moorwood}(1997)}]{moo97b}
{Moorwood} A.F., Mar. 1997, In: Proc. SPIE Vol. 2871, p. 1146-1151, Optical
  Telescopes of Today and Tomorrow, Arne L. Ardeberg; Ed., 1146--1151

\bibitem[{{Moustakas} \& {Somerville}(2002)}]{mou02}
{Moustakas} L.A., {Somerville} R.S., Sep. 2002, \apj, 577, 1

\bibitem[{{Nakajima} \& {Matsuhara}(2001)}]{nak01}
{Nakajima} T., {Matsuhara} H., Feb. 2001, \ao, 40, 514

\bibitem[{{Persi} et~al.(2000){Persi}, {Marenzi}, {Olofsson} et~al.}]{per00}
{Persi} P., {Marenzi} A.R., {Olofsson} G., et~al., May 2000, \aap, 357, 219

\bibitem[{{Pier} \& {Krolik}(1992)}]{pie92}
{Pier} E.A., {Krolik} J.H., Dec. 1992, \apj, 401, 99

\bibitem[{Rees(1998)}]{ree98}
Rees M.J., 1998, Introductory talk at ngst conference, liege, june 1998

\bibitem[{{Rees} \& {Ostriker}(1977)}]{ree77}
{Rees} M.J., {Ostriker} J.P., Jun. 1977, \mnras, 179, 541

\bibitem[{{Rosati} et~al.(2002){Rosati}, {Tozzi}, {Giacconi} et~al.}]{ros02b}
{Rosati} P., {Tozzi} P., {Giacconi} R., et~al., Feb. 2002, \apj, 566, 667

\bibitem[{R{\"o}ttgering et~al.(2000)R{\"o}ttgering, Granato, Guiderdoni, \&
  Rudnick}]{rot00a}
R{\"o}ttgering H., Granato G., Guiderdoni B., Rudnick G., 2000, In: Lena P.J.,
  Quirrenbach A. (eds.) Proceedings of the SPIE conference Interferometry in
  Optical Astronomy, Vol. 4006, 742

\bibitem[{{Rudnick} et~al.(2001){Rudnick}, {Franx}, {Rix} et~al.}]{rud01}
{Rudnick} G., {Franx} M., {Rix} H., et~al., Nov. 2001, \aj, 122, 2205

\bibitem[{{Rudnick} et~al.(2003){Rudnick}, {Rix}, {Franx} et~al.}]{rud03}
{Rudnick} G., {Rix} H., {Franx} M., et~al., Jul. 2003, ArXiv Astrophysics
  e-prints

\bibitem[{{Shu} et~al.(1987){Shu}, {Adams}, \& {Lizano}}]{shu87}
{Shu} F.H., {Adams} F.C., {Lizano} S., 1987, \araa, 25, 23

\bibitem[{{Somerville} et~al.(2001){Somerville}, {Primack}, \& {Faber}}]{som01}
{Somerville} R.S., {Primack} J.R., {Faber} S.M., Jan. 2001, \mnras, 320, 504

\bibitem[{{Steidel} et~al.(1999){Steidel}, {Adelberger}, {Giavalisco},
  {Dickinson}, \& {Pettini}}]{ste99a}
{Steidel} C.C., {Adelberger} K.L., {Giavalisco} M., {Dickinson} M., {Pettini}
  M., Jul. 1999, \apj, 519, 1

\bibitem[{{Trujillo} et~al.(2003){Trujillo}, {Rudnick}, {Rix} et~al.}]{tru03}
{Trujillo} I., {Rudnick} G., {Rix} H., et~al., Jul. 2003, ArXiv Astrophysics
  e-prints

\bibitem[{{Urry} \& {Padovani}(1995)}]{urr95}
{Urry} C.M., {Padovani} P., Sep. 1995, \pasp, 107, 803

\bibitem[{{van Dishoeck}(2000)}]{dis00}
{van Dishoeck} E.F., 2000, In: ASP Conf. Ser. 207: Next Generation Space
  Telescope Science and Technology, 85

\bibitem[{{Venemans} et~al.(2002){Venemans}, {Kurk}, {Miley} et~al.}]{ven02}
{Venemans} B.P., {Kurk} J.D., {Miley} G.K., et~al., Apr. 2002, \apjl, 569, L11

\bibitem[{{White} \& {Frenk}(1991)}]{whi91}
{White} S.D.M., {Frenk} C.S., Sep. 1991, \apj, 379, 52

\bibitem[{{White} \& {Rees}(1978)}]{whi78}
{White} S.D.M., {Rees} M.J., May 1978, \mnras, 183, 341

\bibitem[{{Wolf} \& {Henning}(1999)}]{wol99}
{Wolf} S., {Henning} T., Jan. 1999, \aap, 341, 675

\end{thebibliography}

\end{document}